\documentclass[
preprintnumbers,
preprint,
a4paper,
showpacs, 
amsmath, 
amssymb, 
floatfix,
nofootinbib,
]{revtex4}

\usepackage[dvips]{graphicx}
\begin{document}

%<<<<<<<<<<<<< TITLE >>>>>>>>>>>>>>>%
\title{Collision of high-energy closed strings: \\
Formation of a ring-like apparent horizon}

%<<<<<<<<<<<<< AUTHOR 1 >>>>>>>>>>>>>>>%
\author{Hirotaka Yoshino}

%<<<<<<<<<<<<< ADDRESS 1 >>>>>>>>>>>>>>>%
\affiliation{Department of Physics, University of Alberta, 
Edmonton, Alberta, Canada T6G 2G7}

%<<<<<<<<<<<<< AUTHOR 2 >>>>>>>>>>>>>>>%
\author{Tetsuya Shiromizu}

%<<<<<<<<<<<<< ADDRESS 2 >>>>>>>>>>>>>>>%
\affiliation{Department of Physics, Tokyo Institute of Technology, 
Tokyo 152-8551, Japan}

%<<<<<<<<<<<<< PREPRINT NO >>>>>>>>>>>>>>>%
\preprint{Alberta-Thy-04-07}

%<<<<<<<<<<<<< DATE >>>>>>>>>>>>>>>%
\date{June 30, 2007}

%
%======================================%
%<<<<<<<<<<<<< ABSTRACT >>>>>>>>>>>>>>>%
%======================================%
%
\begin{abstract}
We study collisions of two high-energy closed strings
in the framework of $D$-dimensional general relativity. 
The model of a high-energy closed string is introduced
as a {\it pp}-wave generated by a ring-shaped source
with the radius $R$. At the instant of the collision,
the positions of two strings are assumed to coincide precisely.
In this setup, we study the formation of two kinds of apparent horizons (AHs):
the AH of topology $S^{D-2}$ (the black hole AH) and the AH of
topology $S^1\times S^{D-3}$ (the black ring AH). 
These two AHs are solved numerically and the conditions for the formation 
of the two AHs are clarified in terms of the ring radius $R$.
Specifically, we demonstrate that the black ring AH
forms for sufficiently large $R$. 
The effects of an impact parameter and the relative orientation of incoming strings
in more general cases are briefly discussed. 
\end{abstract}

%<<<<<<<<<<<<< PACS NUMBER >>>>>>>>>>>>>>>%
\maketitle

%
%======================================%
%<<<<<<<<<<<< SECTION I  >>>>>>>>>>>>>>%
%======================================%
%
\section{Introduction}

The trans-Planckian collisions of particles
attract renewed interests motivated
by the scenarios of large extra 
dimensions \cite{ADD98,RS99}.
In these scenarios, the gravity becomes higher dimensional
at microscopic scale and the Planck energy could be as low 
as TeV. This indicates the possibility of the observation
of quantum gravity phenomena at near-future accelerators.
There are two ways to approach this subject. 
One is the string theory, in which one can
study the regime where the string length 
is important but gravity is not so strong
 (see \cite{DE01,V04,GGM07} for recent studies).
Since the regime of strong gravity cannot be investigated 
by the string theory, the theory of general relativity is
often used as the alternative approach. 
In this approach, the intermediate
state is expected to be a black hole. See \cite{BF99,DL01,GT02}
for outlines of the expected phenomena of tiny black holes at accelerators.

To study the process of the black hole production
in general relativity, 
the apparent horizon (AH) is a very useful tool,
since formation of an AH implies the existence of the event horizon
outside of it.
In \cite{EG02}, the grazing collision of high-energy particles
was studied using the Aichelburg-Sexl particle model~\cite{AS71}, 
and the analytic formula for the AH 
was obtained in the four-dimensional case.
The numerical code for solving the AHs of this system 
in the higher-dimensional cases 
was developed by one of us and Nambu~\cite{YN03},
and the results of \cite{YN03} were further 
improved by adopting a better slice~\cite{YR05}.

Since the Aichelburg-Sexl particle is a simplified model 
of a particle, several efforts have been made in 
order to take into account the potentially important effects.
The effect of electric charge was studied in \cite{YM06}, 
of which results were used to improve the estimate
of the black hole production rate at accelerators \cite{G06}.
In \cite{R04, GR04}, the validity of the general relativistic model
was examined and the importance of wavepacket effects
was pointed out. 
In a recent paper, the effects of the particle spin 
and of the wavepacket were studied in \cite{YZF07}
using the gyraton model \cite{FF05,FIZ05}. 
See also \cite{KV02, KN05} for trials to incorporate
the effects from the string theory.

\begin{figure}[tb]
\centering
{
\includegraphics[width=0.25\textwidth]{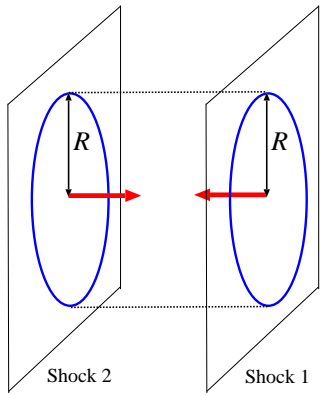}
}
\caption{The system that we study in this paper. Two shock 
gravitational fields generated by high-energy closed strings with
the radius $R$ collide in a $D$-dimensional spacetime.
The positions of the two strings coincide at the instant of the collision.
}
\label{setup}
\end{figure}

In this paper, we study the collisions of high-energy extended objects
in the framework of general relativity. 
Specifically, we consider the system of two ``closed strings''
whose gravitational field is of the {\it pp}-wave type.
Figure~\ref{setup} shows the configuration of
the system that we study in this paper. 
The two gravitational shocks collide in a $D$-dimensional spacetime.
Each shock field is generated by a ring-shaped source
whose radius is $R$. The positions of the two strings
exactly coincide at the instant of the collision.
In this setup, formation of AHs is investigated.

%===========<Picture>============%
%
%
%=================================%

This study has two meanings.
First, it is expected that
if the length of strings 
is taken into account,
the black hole formation in high-energy collisions 
becomes more difficult
compared to the cases of pointlike particles (e.g., \cite{GGM07,KV02,H04,Y04}).
By considering the collisions of extended objects, 
we would be able to incorporate some of effects 
of the strings on the black hole
formation. Such an approach was already done
in \cite{KV02}. In that paper, the collisions of uniform disks
were studied. The reason for this setup comes from
the expectation that the strings behave as quasi-homogeneous 
beams \cite{ACV89}. However, since how high-energy strings couple
to gravity is still an open problem, 
it is meaningful to study the collisions of high-energy shocks with 
one-dimensional sources.
Furthermore, this work will clarify how
the results in \cite{KV02} depend on the energy
density distributions of incoming objects.

Next, we can provide a possible scenario
of producing a black ring as a result of the gravitational collapse.
The black ring \cite{ER02} is the solution of the asymptotically flat 
five-dimensional spacetime with the event horizon
of topology $S^1\times S^2$.
Although the black ring solution has been found
only in the five-dimensional case, there are strong
indications for the existence of $(D\ge 6)$-dimensional 
black ring solutions \cite{IN02},
whose event horizon will have the topology $S^1\times S^{D-3}$.
Unfortunately the black rings are expected to be unstable
(see, e.g., Sec. I of \cite{YN04}). However, since they would be
able to exist as intermediate states of gravitational collapses,
it is of interest to provide scenarios of producing AHs of topology 
$S^1\times S^{D-3}$ (say, black ring AHs).
The possibility of producing black rings in collisions of two high-energy particles
was discussed in \cite{IOP03}. 
In \cite{YN04}, it was clarified that multi-particle systems
can make the black ring AHs. In this paper, we demonstrate that
collisions of closed strings also lead to the formation of the black ring AHs.
We study both the black ring AH and 
the AH of topology $S^{D-2}$ (say, black hole AH), 
and clarify the conditions for the formation of
the two AHs in terms of the ring radius $R$.

This paper is organized as follows. In the next section,
we introduce the model of a high-energy closed string
and set up the collision. In Sec. III,
we show the AH equation and the boundary conditions,
and demonstrate existence of the black ring AH
for large $R$. The numerical methods for 
solving the two AHs are also explained. 
Then the numerical results are shown in Sec. IV.
Section V is devoted to discussions on more general collisions 
and implications for trans-Planckian collisions in
the scenarios of large extra dimensions.
In Appendix A, formulas necessary for calculating the metrics 
of high-energy closed strings are presented.

%
%======================================%
%<<<<<<<<<<<< SECTION II  >>>>>>>>>>>>>>%
%======================================%
%
\section{System setup}

In this section, we introduce the model of a
high-energy closed string and set up the collision.

%
%======================================%
%<<<<<<<<<< subsection 2.1 >>>>>>>>>>>>%
%======================================%
%
\subsection{Model of a high-energy closed string}

In order to introduce the model of a high-energy closed string
in a $D$-dimensional spacetime,
we assume the following metric of the {\it pp}-wave form:
%===========<Equation>============%
%
\begin{equation}
ds^2=-d\bar{u}d\bar{v}+\sum_{i=1}^{D-2}d\bar{x}_i^2+\Phi(\bar{x}_i)\delta(\bar{u})d\bar{u}^2.
\label{pp-wave-metric}
\end{equation}
%
%=================================%
The spacetime is flat except at $\bar{u}=0$. There exists a gravitational 
shock wave at $\bar{u}=0$, which propagates at the speed of light.
The properties of the shock gravitational field is characterized by the 
function $\Phi(\bar{x}_i)$, which we call the shock potential hereafter. 
The nonzero component of the
energy-momentum tensor of this spacetime has the form
%===========<Equation>============%
%
\begin{equation}
T_{\bar{u}\bar{u}}=\hat{\rho}(\bar{x}_i)\delta(\bar{u}).
\end{equation}
Here, $\hat{\rho}(\bar{x}_i)$ indicates the energy density in the shock.
The Einstein equation is
%===========<Equation>============%
%
\begin{equation}
\bar{\nabla}^2\Phi=-16\pi G\hat{\rho}(\bar{x}_i)
\end{equation}
%
%=================================%
with the $D$-dimensional gravitational constant $G$
and the flat space Laplacian $\bar{\nabla}^2$ in the coordinates $\bar{x}_i$.
The Aichelburg-Sexl particle \cite{AS71} has the shock potential
%===========<Equation>============%
%
\begin{equation}
\Phi=
\begin{cases}
-8Gp\log\bar{r}, & (D=4);\\
\displaystyle \frac{16\pi Gp}{(D-4)\Omega_{D-3}\bar{r}^{D-4}}, & (D\ge 5).
\end{cases}
\label{AS-potential}
\end{equation}
%
%=================================%
Here, $\bar{r}:=\sqrt{\sum_i\bar{x}_i^2}$ and
$\Omega_{D-3}$ is the area of the $(D-3)$-dimensional unit sphere. 
Since the energy density of this AS particle is 
$\hat{\rho}=p\delta^{D-2}(\bar{x}_i)$, 
the Aichelburg-Sexl particle represents a point 
particle with the energy $p$.

We would like to generalize the Aichelburg-Sexl particle
to high-energy closed strings. Let us introduce coordinates
$(\bar{x}, \bar{y}, \bar{z}_i)$ by
%===========<Equation>============%
%
\begin{equation}
\bar{x}=\bar{x}_1,~~ \bar{y}=\bar{x}_2, ~~\bar{z}_i=\bar{x}_{i+2},
\end{equation}
%
%=================================%
with $i=1,...,D-4$, 
and put the energy source on the $(\bar{x}, \bar{y})$-plane 
in the shape of a ring with the radius $R$.
If we introduce coordinates $(\bar{W}, \bar{\phi})$ by
%===========<Equation>============%
%
\begin{equation}
\bar{x}=\bar{W}\cos\bar{\phi},~~\bar{y}=\bar{W}\sin\bar{\phi},
\label{W-coordinate}
\end{equation} 
%
%=================================%
the energy density we assume is
%===========<Equation>============%
%
\begin{equation}
\hat{\rho}=p\frac{\delta(\bar{W}-R)}{2\pi\bar{W}}\delta^{D-4}(\bar{z}_i).
\end{equation}
For $D\ge 5$, the shock potential for this energy density is given by
%===========<Equation>============%
%
\begin{equation}
\Phi=\frac{16\pi Gp}{(D-4)\Omega_{D-3}}
\int_0^{2\pi} 
\frac{d\zeta/2\pi}
{\left[(\bar{x}-R\cos\zeta)^2+(\bar{y}-R\sin\zeta)^2+\bar{Z}^2\right]^{(D-4)/2}}
\end{equation}
%
%=================================%
with
%===========<Equation>============%
%
\begin{equation}
\bar{Z}:=\sqrt{\sum \bar{z}_i^2}.
\label{Z-coordinate}
\end{equation}

Hereafter, we adopt 
%===========<Equation>============%
%
\begin{equation}
r_0:=\left(\frac{8\pi Gp}{\Omega_{D-3}}\right)^{1/(D-3)}
\end{equation}
%
%=================================%
as the unit of the length. For $D=5$--$11$, the value of $r_0$ 
is related to the gravitational radius $r_h(2p)$ of the system 
as $r_0\simeq 1.1r_h(2p)$.
In order to calculate the black hole AH, it is useful to introduce 
the spherical-polar coordinates $(\bar{r}, \bar{\theta}, \bar{\phi})$:
%===========<Equation>============%
%
\begin{align}
\bar{x}=\bar{r}\sin\bar{\theta}\cos\bar{\phi}, ~~
\bar{y}=\bar{r}\sin\bar{\theta}\sin\bar{\phi}, ~~
\bar{Z}=\bar{r}\cos\bar{\theta}.
\label{spherical-polar}
\end{align}
%
%=================================%
In these coordinates, the shock potential is written as
%===========<Equation>============%
%
\begin{equation}
\Phi=\frac{I_D}{\pi (D-4)},
\end{equation}
%
%=================================%
%===========<Equation>============%
%
\begin{equation}
I_D:=
\int_0^{2\pi}\frac{d\zeta}{(a-b\cos\zeta)^{(D-4)/2}},
\label{integral}
\end{equation}
%
%=================================%
with
%===========<Equation>============%
%
\begin{equation}
a=\bar{r}^2+R^2,~~b=2R\bar{r}\sin\bar{\theta}.
\end{equation}
%
%=================================%
The results of the integration of $I_D$ for $D=5$--$11$ are summarized in
the appendix. They are given in terms of elementary functions
for even $D$ while in terms of 
the complete elliptic integrals for odd $D$.

The above formula holds only for $D\ge 5$. For four dimensions,
the shock potential is
%===========<Equation>============%
%
\begin{equation}
\Phi=\begin{cases}
-2\ln \bar{r}, & (\bar{r}\ge R);\\
-2\ln R, & (\bar{r}\le R),
\end{cases}
\end{equation}
%
%=================================%
which has the same form as that of the Aichelburg-Sexl
particle outside of the ring. Therefore, in the case $D=4$,
we can easily see what happens in the head-on collision
of closed strings using the results of \cite{EG02}.
The black hole AH forms if and only if $R<1$.
We also find that
the black ring AH does not form for any values of $R$.
This is consistent with the theorem for the topology of AHs
in four dimensions \cite{H72}.
Hereafter we will consider higher dimensional cases with $D\ge 5$.

We note that the above metrics 
are special cases of ``string gyratons'' proposed in \cite{FL06}.
In that paper, solutions of the gravitational field with the rank 3 antisymmetric
Kalb-Ramon field generated by stringlike sources were obtained,
and our metrics are the zero Kalb-Ramon charge cases of those solutions.
We also note that our metric for $D=5$ coincides
with the one obtained by taking the lightlike limit of
the boosted black ring solution \cite{OKP05}.
As we mentioned in Sec. I, it is expected that the black ring solutions
exist also for $D\ge 6$. Hence we conjecture
that the lightlike boosts of such black ring solutions 
will give the above metrics after discovery of exact
solutions for $D\ge 6$.

Let us examine the behavior of the shock potential $\Phi$
in the neighborhood of the ring. For this purpose, we introduce 
new coordinates $(\bar{\rho}, \bar{\xi}, \bar{\phi})$ by
%===========<Equation>============%
%
\begin{equation}
\bar{x}=(R+\bar{\rho}\cos\bar{\xi})\cos\bar{\phi},~~
\bar{y}=(R+\bar{\rho}\cos\bar{\xi})\sin\bar{\phi},~~
\bar{Z}=\bar{\rho}\sin\bar{\xi}.
\label{ring-coordinates}
\end{equation}
%
%=================================%
These coordinates are also useful for studying the black ring AH.
In these coordinates, the ring is located at $\bar{\rho}=0$.
In the regime $\bar{\rho}/R\ll 1$, the shock potential behaves as
%===========<Equation>============%
%
\begin{equation}
\Phi\simeq 
\begin{cases}
\displaystyle -\frac{2}{\pi R}\log \left(\frac{\bar{\rho}}{2R}\right), & (D=5);\\
\displaystyle \frac{\alpha_D}{R\bar{\rho}^{D-5}},& (D\ge 6),
\end{cases}
\label{potential-largeR}
\end{equation}
%
%=================================%
with
%===========<Equation>============%
%
\begin{equation}
\alpha_D:=\frac{\Gamma((D-5)/2)}{2\sqrt{\pi}\Gamma(D/2-1)}.
\end{equation}
%
%=================================%
These formulas will be used later to
show the existence of the black ring AH in the case $R\gg 1$.

%
%======================================%
%<<<<<<<<<< subsection 2.2 >>>>>>>>>>>>%
%======================================%
%
\subsection{Continuous and smooth coordinates}

Because the delta function in the metric \eqref{pp-wave-metric}
indicates the discontinuity of the coordinates 
$(\bar{u}, \bar{v}, \bar{x}_i)$ at $\bar{u}=0$, 
it is necessary to introduce the continuous and
smooth coordinates $(u, v, x_i)$ in order to set up the collision.
Such a coordinate transformation is given in \cite{EG02}
as 
%===========<Equation>============%
%
\begin{align}
\bar{u}&=u,\\
\bar{v}&=v+\Phi \theta(u)+\frac14u\theta(u)(\nabla\Phi)^2,\\
\bar{x}_i&=x_i+\frac12 u\nabla_i\Phi(x)\theta(u).
\end{align}
%
%=================================%
Here, $\theta(u)$ denotes the Heaviside step function.
In these coordinates, the line $v,x_i=\mathrm{const.}$
is the null geodesic and $u$ is its affine parameter.
The metric is written as
%===========<Equation>============%
%
\begin{equation}
ds^2=-dudv+H_{ik}H_{jk}dx^idx^j,
\end{equation}
%
%=================================%
%===========<Equation>============%
%
\begin{equation}
H_{ij}=\delta_{ij}+\frac12\nabla_i\nabla_j\Phi(\boldsymbol{x})u\theta(u).
\end{equation}
%
%=================================%

%
%======================================%
%<<<<<<<<<< subsection 2.3 >>>>>>>>>>>>%
%======================================%
%
\subsection{Setup of the collision}  

%===========<Picture>============%
%
\begin{figure}[tb]
\centering
{
\includegraphics[width=0.25\textwidth]{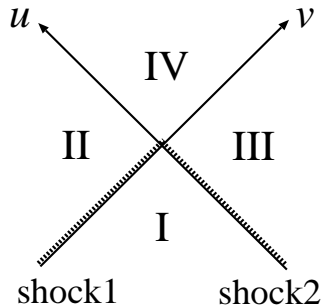}
}
\caption{ Schematic structure of the spacetime for the system of two high-energy strings.
The metric in the regions I, II, and III can be written down. We study the AHs on the slice
indicated by a dotted line.}
\label{collision}
\end{figure}
%
%=================================%

Now we set up the collision of two high-energy closed strings.
Figure \ref{collision} shows the schematic picture of the spacetime structure.
The spacetime is divided into regions I: $(u\le 0, v\le 0)$, II: $(u\ge 0, v\le 0)$,
III: $(u\le 0, v\ge 0)$, and IV: $(u\ge 0, v\ge 0)$. Because the strings propagate
at the speed of light, they do not interact before the collision. Therefore
we can write down the metric in regions I, II, and III just by combining
the two metrics: 
%===========<Equation>============%
%
\begin{equation}
ds^2=-dudv+\left[H_{ik}^{(1)}H_{jk}^{(1)}+H_{ik}^{(2)}H_{jk}^{(2)}-\delta_{ij}\right]dx^idx^j
\end{equation}
%
%=================================%
%===========<Equation>============%
%
\begin{align}
H_{ij}^{(1)}&=\delta_{ij}+\frac12\nabla_i\nabla_j\Phi(\boldsymbol{x})u\theta(u),\\
H_{ij}^{(2)}&=\delta_{ij}+\frac12\nabla_i\nabla_j\Phi(\boldsymbol{x})v\theta(v).
\end{align}
%
%=================================%
Since we use the same formula of the shock potential for
two incoming strings, their positions exactly coincide 
at the instant of the collision.

Region IV is the interaction region and its spacetime structure
is not known. However, it is still possible to confirm the black hole formation
by studying an AH on some slice in regions I, II, and III.
In the next section, we explain how to find the AHs 
on the slice $u\le 0=v$ and $v\le 0=u$ 
(the dotted line in Fig. \ref{collision}). 

%
%======================================%
%<<<<<<<<<<<< SECTION III  >>>>>>>>>>>>>>%
%======================================%
%
\section{Finding apparent horizons}

In this section, we show the equations and the boundary conditions 
for the black hole AH and the black ring AH on the slice $u\le 0=v$
and $v\le 0=u$. These AH equations can be solved analytically 
in the two cases $R=0$ and $R\gg 1$. 
Then, the numerical methods for solving the two AHs 
are presented.

%
%======================================%
%<<<<<<<<<< subsection 3.1 >>>>>>>>>>>>%
%======================================%
%
\subsection{The equation and the boundary conditions}

The AH is defined as a $(D-2)$-dimensional
surface whose outgoing null geodesic congruence has zero expansion.
Let the AH be given by the union of two 
surfaces $S_1$: $v=-\varPsi_1(\boldsymbol{x})$ 
in $v\le 0=u$ and $S_2$: $u=-\varPsi_2(\boldsymbol{x})$ in $u\le 0=v$.
Here, the surfaces $S_{1,2}$ are connected
on a common boundary $C$ in $u=v=0$.
The equation and the boundary conditions for $\varPsi_{1,2}$
were derived in \cite{EG02} (see also \cite{KV02, KN05}).
Here we just comment on the results.
The AH equation is found as $\nabla^2(\varPsi_{1,2}-\Phi_{1,2})=0$,
by imposing the expansion of the null geodesic
congruence of each surface to be zero.
The boundary conditions are
 $\varPsi_{1,2}=0$ and 
$\nabla\varPsi_1\cdot\nabla\varPsi_2=4$ on $C$,    
which come from the continuity of the surface and the null tangent vectors,
respectively.
There are two boundary conditions because the boundary
$C$ itself is an unknown surface to be solved.

%===========<Picture>============%
%
\begin{figure}[tb]
\centering
{
\includegraphics[width=0.3\textwidth]{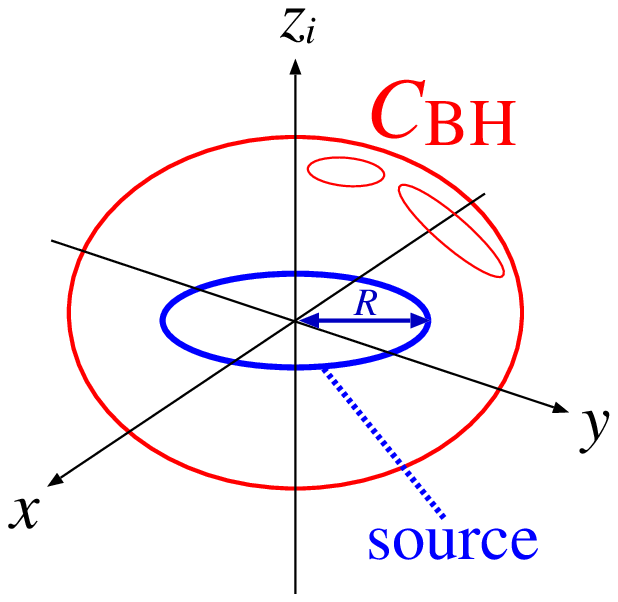}\hspace{5mm}
\includegraphics[width=0.3\textwidth]{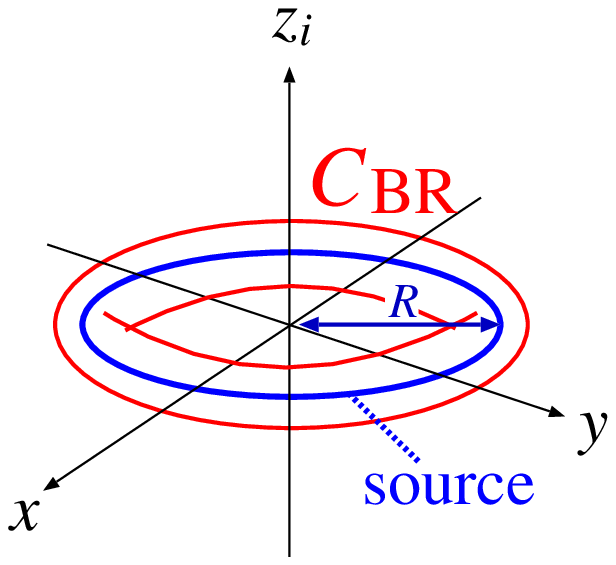}
}
\caption{ Schematic pictures of the boundary 
$C_{\rm BH}$ for the black hole AH (left)
and the boundary $C_{\rm BR}$
for the black ring AH (right).}
\label{boundaries}
\end{figure}
%
%=================================%

As the two incoming strings are identical in our setup, 
we can simply put $\varPsi:=\varPsi_1=\varPsi_2$.
In numerical calculations, it is convenient to introduce
a function $h:=\varPsi-\Phi$. In terms of $h$, the AH equation and the
boundary conditions become
%===========<Equation>============%
%
\begin{equation}
\nabla^2h=0,~~~~~~~\textrm{within}~~C,
\end{equation}
%
%=================================%
%===========<Equation>============%
%
\begin{align}
h=\Phi ~~~\textrm{and}~~~[\nabla(\Phi-h)]^2=4,~~~~~~~\textrm{on}~~C.
\end{align}
%
%=================================%
These equation and boundary conditions can be applied
to both the black hole AH and the black ring AH. 
However, we have to choose the boundary $C$ appropriately in each case.
Figure~\ref{boundaries} shows schematic shapes of the boundaries
$C_{\rm BH}$ and $C_{\rm BR}$ for the black hole AH and 
the black ring AH, respectively.
In the case of the black hole AH, we require the boundary $C_{\rm BH}$
to have the topology $S^{D-3}$ so that the AH can have the topology $S^{D-2}$.
On the other hand, 
the boundary $C_{\rm BR}$ of the black ring AH
is assumed to have the topology $S^1\times S^{D-4}$ 
in order that the AH can have the topology $S^1\times S^{D-3}$.

%
%======================================%
%<<<<<<<<<< subsection 3.2 >>>>>>>>>>>>%
%======================================%
%
\subsection{The cases $R=0$ and $R\gg 1$}

Let us consider the black hole AH in the case $R=0$.
In this case, the system is reduced to the
head-on collision of Aichelburg-Sexl particles.
Using the spherical-polar coordinates $(r,\theta,\phi)$
introduced in Eq.~\eqref{spherical-polar}
(here bars are omitted since we are now working in the continuous coordinates),
the solutions of $h(r,\theta)$ and $C_{\rm BH}$
are given by $h(r,\theta)=\frac{2}{D-4}$ and $r=1$, respectively.

Next we consider the limit $R\gg 1$. 
In this case, we can show the existence of the black ring AH 
as follows.
In the neighborhood of the ring, the shock potential $\Phi$
is approximated as Eq.~\eqref{potential-largeR}.
Then, the solutions of $h(r,\xi)$ and $C_{\rm BR}$
are given by $h(r,\xi)=\Phi(\rho_h)$ and $\rho=\rho_h$, respectively,
where
%===========<Equation>============%
%
\begin{equation}
\rho_h=\left(\frac{\beta_D}{R}\right)^{1/(D-4)}.
\label{rhoH}
\end{equation} 
%
%=================================%
Here, $\beta_D:=\frac{D-5}{2}\alpha_D$.
Because $\rho_h$ becomes small as $R$ is increased,
the black ring AH is located near the ring for large $R$.
It is reminded that the above approximation is valid only for $R\gg 1$.

Since we have the solution of the black hole AH for $R=0$,
we can numerically solve it for $R>0$ 
by slowly increasing the value of $R$.
Similarly, since the solution of the black ring AH is known for $R\gg 1$,
the numerical calculation can be done 
by starting at sufficiently large $R$ and gradually
decreasing the value of $R$.
In the following, we briefly explain the numerical
methods for solving the black hole AH
and the black ring AH, one by one.

%
%======================================%
%<<<<<<<<<< subsection 3.3 >>>>>>>>>>>>%
%======================================%
%
\subsection{Numerical method for black hole AHs}

In order to solve the black hole AH, the spherical-polar coordinates
$(r,\theta,\phi)$ are useful.
In these coordinates, the AH equation is
%===========<Equation>============%
%
\begin{equation}
h_{,rr}+\frac{h_{,\theta\theta}}{r^2}
+\frac{D-3}{r}h_{,r}+\frac{\cot\theta-(D-5)\tan\theta}{r^2}
h_{,\theta}=0.
\end{equation}
%
%=================================%
Let the boundary $C_{\rm BH}$ be given by $r=g(\theta)$.
We further perform a coordinate transformation
%===========<Equation>============%
%
\begin{equation}
\tilde{r}:=r/g(\theta),
\end{equation}
%
%=================================%
for numerical convenience. In the coordinates $(\tilde{r},\theta)$, the AH 
equation becomes
%===========<Equation>============%
%
\begin{multline}
\left(1+\frac{g^{\prime 2}}{g^2}\right)h_{,\tilde{r}\tilde{r}}
-2\frac{g^\prime}{g}\frac{h_{,\tilde{r}\theta }}{\tilde{r}}
+\frac{h_{\theta \theta }}{\tilde{r}^2}
+\left[
D-3-\left(\frac{g^{\prime\prime}}{g}-2\frac{g^{\prime 2}}{g^2}\right)
\right]\frac{h_{,\tilde{r}}}{\tilde{r}}
\\
+\frac{\cot\theta -(D-5)\tan\theta }{\tilde{r}^2}
\left(h_{,\theta }-\tilde{r}\frac{g^\prime}{g}h_{,\tilde{r}}\right)
=0.
\end{multline}
%
%=================================%
Then, we can write down the finite difference equations with 
the second order accuracy.

At the coordinate singularity $\tilde{r}=0$, we cannot use the above
equation. Instead, we used the $(W,Z)$ coordinates introduced in 
Eqs. \eqref{W-coordinate} and \eqref{Z-coordinate}.
In these coordinates, the AH equation is
%===========<Equation>============%
%
\begin{equation}
2h_{,WW}+(D-4)h_{,ZZ}=0
\label{WZ-eq}
\end{equation}
%
%=================================%
at $W=Z=0$ (i.e., $\tilde{r}=0$). We wrote down the finite difference
equation of Eq.~\eqref{WZ-eq}, which can be rewritten as the 
equation to determine the value of $h$ at $\tilde{r}=0$
in the original coordinates $(\tilde{r},\theta )$.

In order to solve this problem, we
used the code developed in \cite{YN03}, which 
makes both $h(\tilde{r},\theta )$ and $g(\theta)$
simultaneously converge to the real solutions. 
We used the grid numbers $(50\times 50)$.
By comparing the results with the one
obtained by the doubled grid numbers $(100\times 100)$,
the characteristic numerical error is estimated to be 0.2\%
for $D=5$--$11$.

%
%======================================%
%<<<<<<<<<< subsection 3.4 >>>>>>>>>>>>%
%======================================%
%
\subsection{Numerical method for black ring AHs}

In order to solve the black ring AH, we use the coordinates
$(\rho,\xi,\phi)$ introduced in Eq.~\eqref{ring-coordinates}.
In these coordinates, the AH equation is
%===========<Equation>============%
%
\begin{equation}
h_{,\rho\rho}+\frac{h_{,\xi\xi}}{\rho^2}
+\left(\frac{D-4}{\rho}+\frac{\cos\xi}{R+\rho\cos\xi}\right)h_{,\rho}
+\frac{1}{\rho^2}
\left[(D-5)\cot\xi-\frac{\rho\sin\xi}{R+\rho\cos\xi}\right]h_{,\xi}=0.
\end{equation}
%
%=================================%
Similar to the case of the black hole AH,
we give the boundary $C_{\rm BR}$ by $\rho=f(\xi)$.
Applying a coordinate transformation
%===========<Equation>============%
%
\begin{equation}
\tilde{\rho}:=\rho/f(\xi),
\end{equation}
%
%=================================%
the AH equation becomes
%===========<Equation>============%
%
\begin{multline}
\left(1+\frac{f^{\prime 2}}{f^2}\right)h_{,\tilde{\rho}\tilde{\rho}}
-2\frac{f^\prime}{f}\frac{h_{,\tilde{\rho}\xi}}{\tilde{\rho}}
+\frac{h_{\xi\xi}}{\tilde{\rho}^2}
+\left[
\frac{1}{\tilde{\rho}}\left(
D-4-\frac{f^{\prime\prime}}{f}+2\frac{f^{\prime 2}}{f^2}\right)
+\frac{\cos\xi}{R/f+\tilde{\rho}\cos\xi}
\right]
h_{,\tilde{\rho}}
\\
+\frac{1}{\tilde{\rho}^2}
\left[
(D-5)\cot\xi-\frac{\tilde{\rho}\sin\xi}{R/f+\tilde{\rho}\cos\xi}
\right]
\left(h_{,\xi}-\tilde{\rho}\frac{f^\prime}{f}h_{,\tilde{\rho}}\right)
=0.
\end{multline}
%
%=================================%
Then the finite difference equations with 
the second order accuracy can be written down.

Similar to the case of the black hole AH, a careful treatment
is required at $\tilde{\rho}=0$. We used the coordinates 
$(X,Y):=(\rho\cos\xi,\rho\sin\xi)$, in which the AH equation becomes
%===========<Equation>============%
%
\begin{equation}
h_{,XX}+(D-4)h_{,YY}+\frac{h_{,X}}{R}=0
\end{equation}
%
%=================================%
at $X=Y=0$ (i.e., $\tilde{\rho}=0$). 
Since there is no symmetry with respect to the $Y$ axis in this case,
the discretization becomes complicated.
It is still possible to prepare the finite difference equation with the
second order accuracy and rewrite it 
in terms of the original coordinates $(\tilde{\rho},\xi)$.

We solved $h(\tilde{\rho},\xi)$ and $f(\xi)$
by rewriting the code of \cite{YN03}
using the grid numbers $(50\times 100)$.
By comparing the results with the one obtained by 
the doubled grid numbers $(100\times 200)$,
the characteristic numerical error is estimated to be 0.1--0.3\%
for $D=5$--$11$.

%
%======================================%
%<<<<<<<<<<<< SECTION IV  >>>>>>>>>>>>>>%
%======================================%
%
\section{Numerical results}

In this section, we show the numerical results
for the black hole AH and the black ring AH.
The amounts of energy trapped by the produced 
AHs are also evaluated.
Then we briefly discuss the interpretation of the obtained results.

%
%======================================%
%<<<<<<<<<< subsection 4.1 >>>>>>>>>>>>%
%======================================%
%
\subsection{Black hole AHs}

%===========<Picture>============%
%
\begin{figure}[tb]
\centering
{
\includegraphics[width=0.4\textwidth]{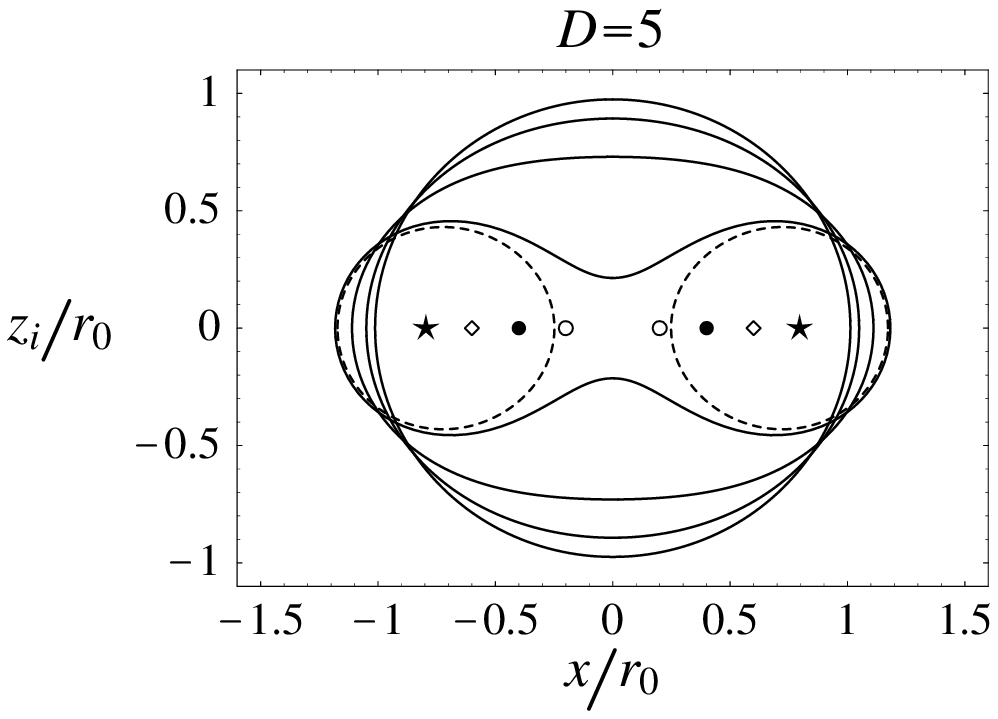}\hspace{5mm}
\includegraphics[width=0.4\textwidth]{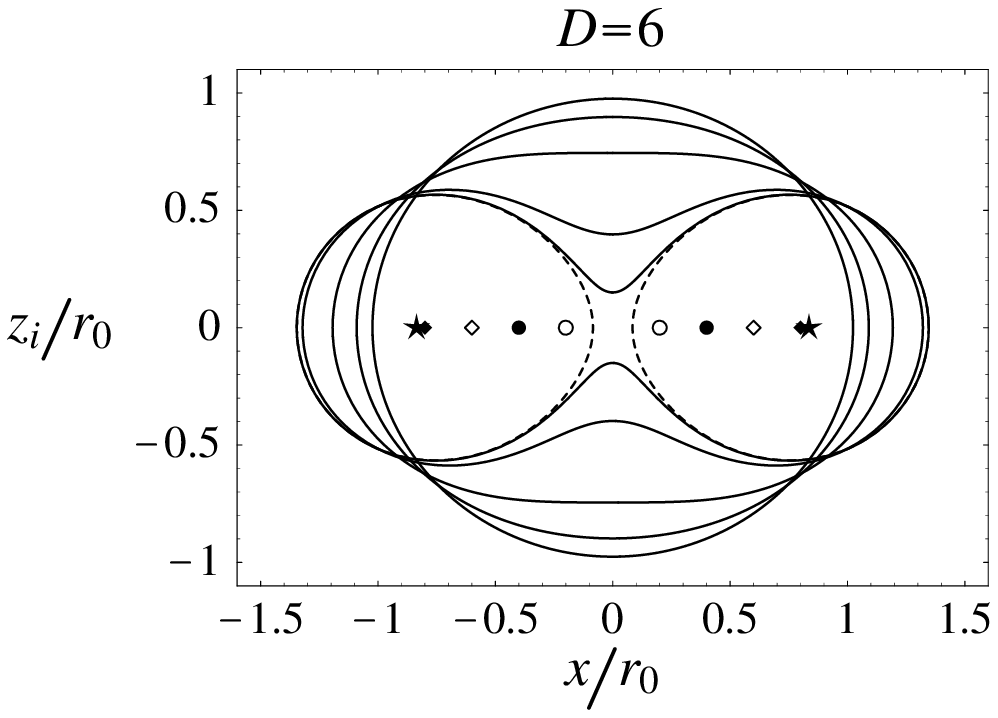}\\
\includegraphics[width=0.4\textwidth]{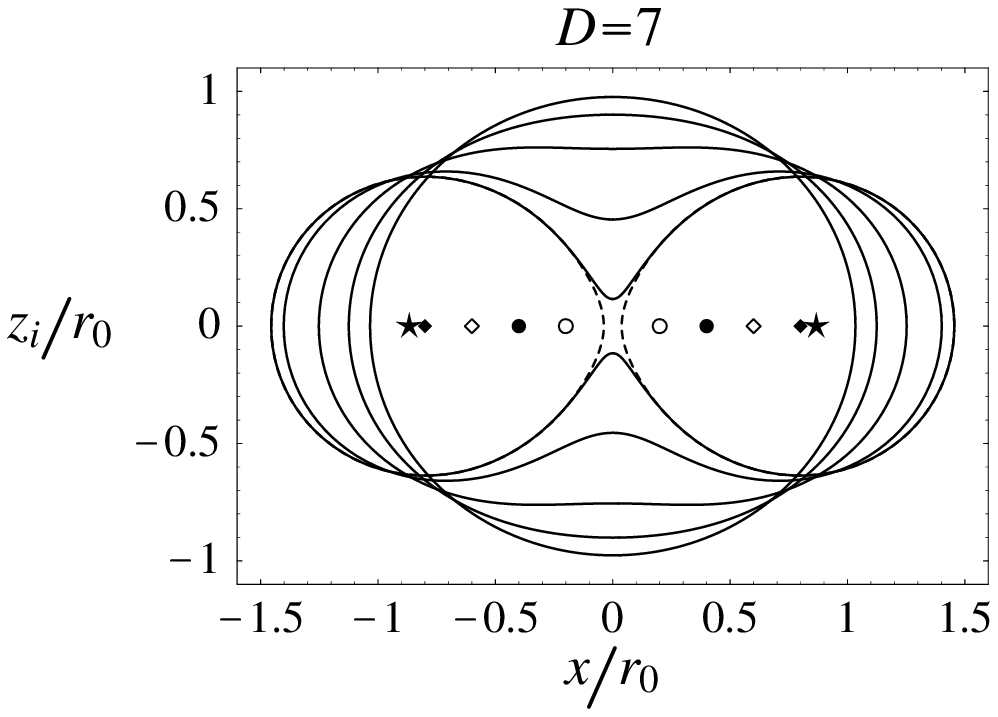}\hspace{5mm}
\includegraphics[width=0.4\textwidth]{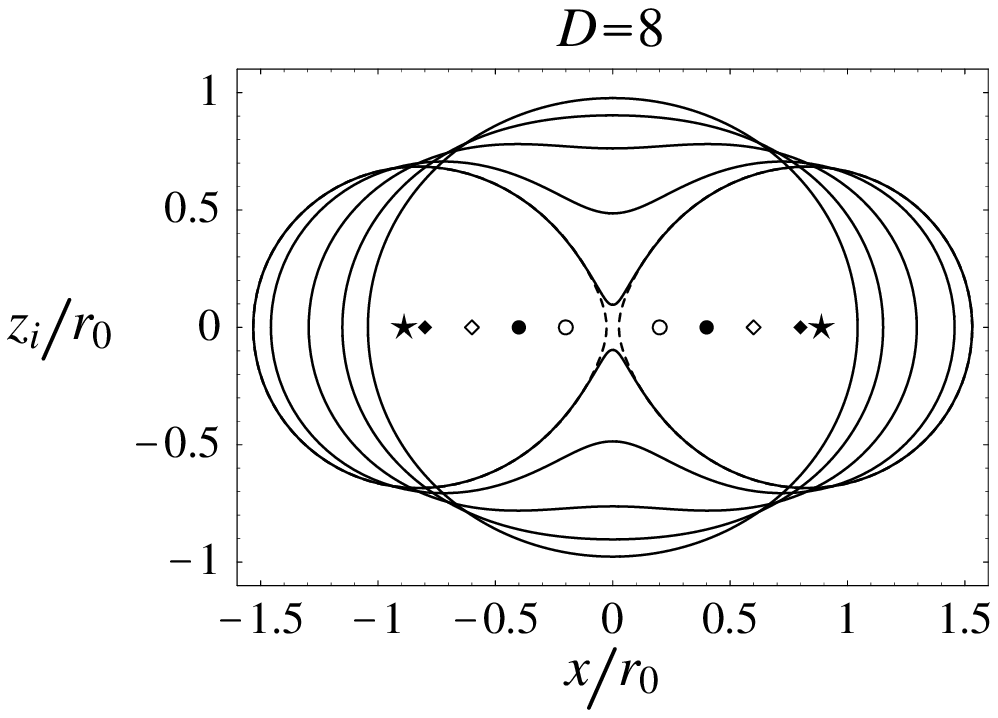}
}
\caption{ Shapes of the boundary $C_{\rm BH}$ (the solid lines)
for the black hole AH on the $(x,z_i)$-plane for $D=5$--$8$. 
The boundary $C_{\rm BH}$ becomes oblate as $R$ is increased.
The values of the ring radius $R/r_0$ are $0.2$($\circ$),
$0.4$($\bullet$), $0.6$($\lozenge$),
$0.8$($\blacklozenge$, shown only for $D=6$--$8$), and
$R_{\text{max}}^{\rm (BH)}/r_0$ ($\star$).
For $R=R_{\text{max}}^{\rm (BH)}$, the black ring AH (the dashed line) 
exists inside of the black hole AH.}
\label{blackholeAHs}
\end{figure}
%
%=================================%

Figure~\ref{blackholeAHs} shows shapes of the boundary $C_{\rm BH}$
of the black hole AH on the $(x,z_i)$-plane for $D=5$--$8$. 
The $y$-axis is suppressed.
As the ring radius $R$ is increased, the black hole AH becomes oblate.
There is some critical value $R_{\text{max}}^{\rm (BH)}$
such that the solution of the black hole AH cannot be found for 
$R>R_{\text{max}}^{\rm (BH)}$. At $R=R_{\text{max}}^{\rm (BH)}$,
the black ring AH also exists inside of the black hole AH.
For large $D$, their shapes agree well except at small $\theta$.

%===========<Picture>============%
%
\begin{figure}[tb]
\centering
{
\includegraphics[width=0.5\textwidth]{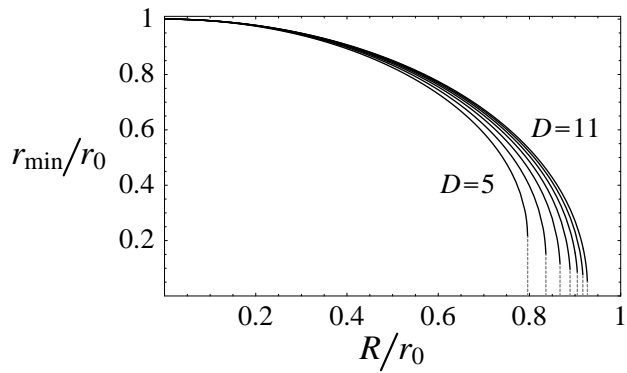}
}
\caption{ The relation between the ring radius $R$
and the minimum radius of the $C_{\rm BH}$,
i.e. $r_{\rm min}:=g(0)$, for $D=5$--$11$. 
The value of $dr_{\rm min}/dR$ becomes $-\infty$ 
at $R=R_{\rm max}^{\rm (BH)}$. }
\label{minimumradius}
\end{figure}
%
%=================================%

The function $g(\theta)$ that specifies the location of $C_{\rm BH}$
takes a minimum value at $\theta=0$.
In Fig. \ref{minimumradius},
we plot this ``minimum radius'' $r_{\rm min}:=g(0)$ 
as a function of $R$ for each $D$. The minimum radius $r_{\rm min}$
is a monotonically decreasing function of $R$, and its gradient becomes
$-\infty$ at $R=R_{\text{max}}^{\rm (BH)}$. 
The values of $R_{\text{max}}^{\rm (BH)}$ are 
somewhat smaller than $r_0$, 
as summarized in Table~\ref{table1}.

%
%======================================%
%<<<<<<<<<< subsection 4.2 >>>>>>>>>>>>%
%======================================%
%
\subsection{Black ring AHs}

%===========<Picture>============%
%
\begin{figure}[tb]
\centering
{
\includegraphics[width=0.5\textwidth]{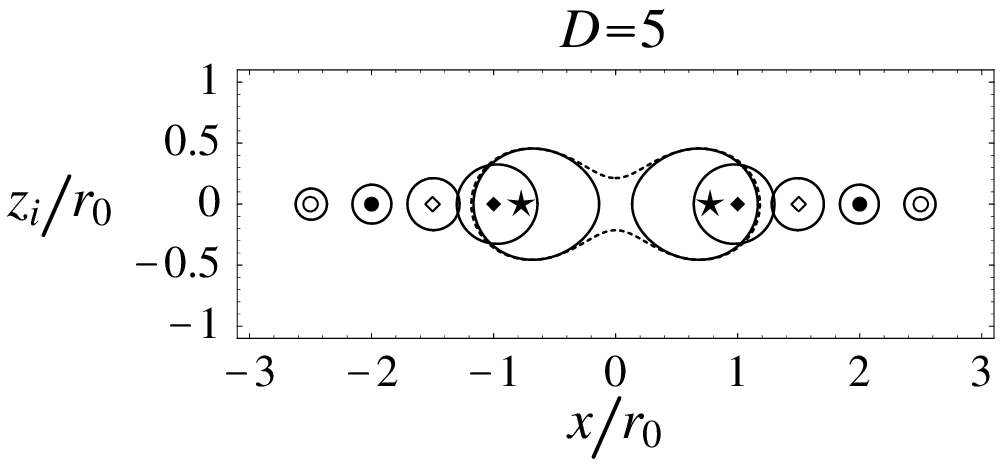}\\
\includegraphics[width=0.5\textwidth]{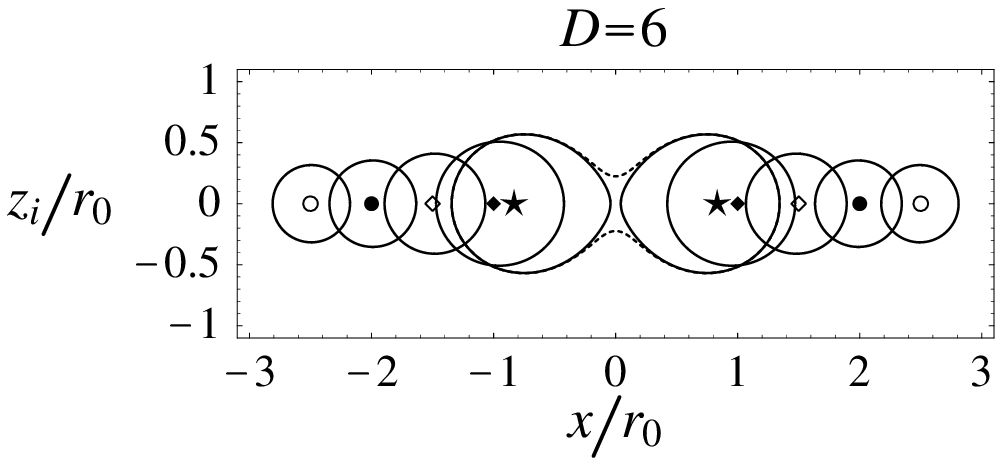}\\
\includegraphics[width=0.5\textwidth]{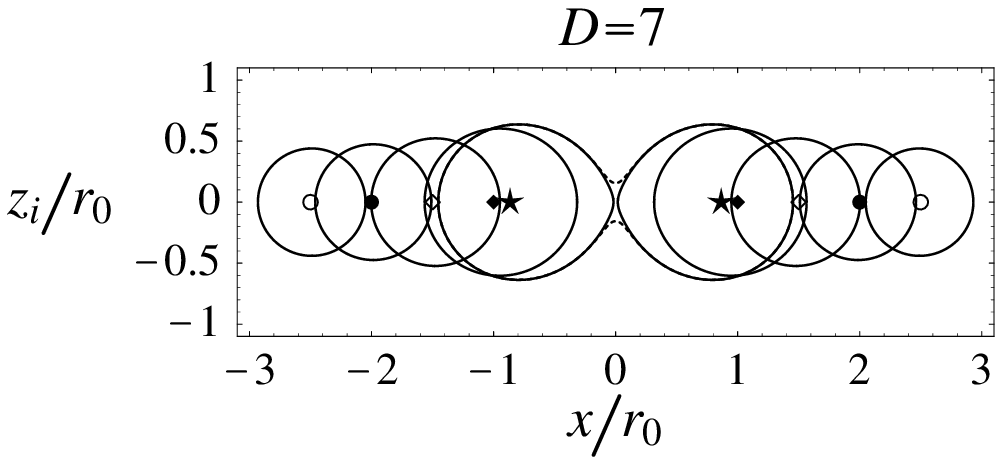}\\
\includegraphics[width=0.5\textwidth]{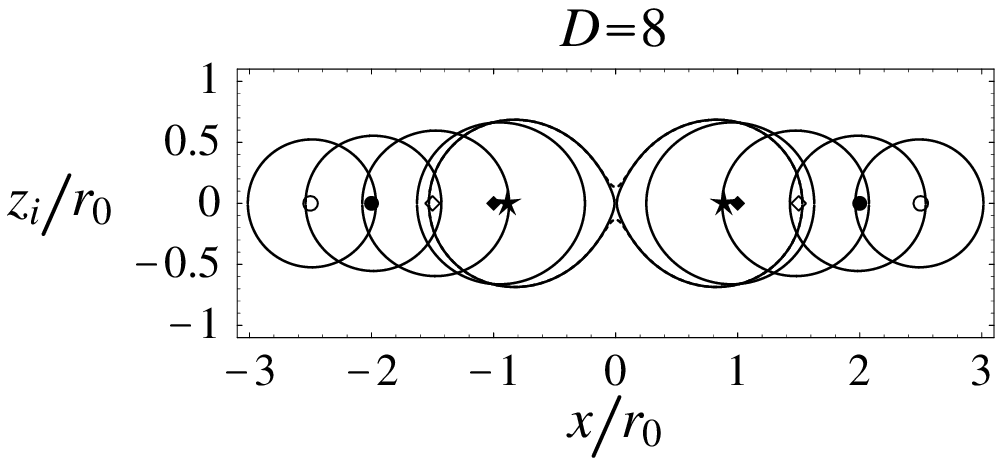}
}
\caption{ Shapes of the boundary $C_{\rm BR}$ (the solid lines)
for the black ring AH on the $(x,z_i)$-plane for $D=5$--$8$.
The values of the ring radius $R/r_0$ are $2.5$($\circ$),
$2.0$($\bullet$), $1.5$($\lozenge$), $1.0$($\blacklozenge$), and
$R_{\text{min}}^{\rm (BR)}/r_0$ ($\star$).
For $R=R_{\text{min}}^{\rm (BR)}$, the black hole AH (the dashed line) 
exists outside of the black ring AH.}
\label{blackringAHs}
\end{figure}
%
%=================================%

Figure~\ref{blackringAHs} shows shapes of the boundary $C_{\rm BR}$
of the black ring AH on the $(x,z_i)$-plane for $D=5$--$8$. 
Since the boundary $C_{\rm BR}$ has the topology $S^1\times S^{D-4}$,
it has two characteristic scales: the radius of the $S^1$ circle
and the radius of the $S^{D-4}$ sphere.
For large $R$,
the radius of the $S^{D-4}$ sphere is approximately equal to $\rho_h$
defined in Eq.~\eqref{rhoH}, and becomes large as
$R$ is decreased.
The black ring AH disappears at some critical value 
$R=R_{\text{min}}^{\rm (BR)}$.
At $R=R_{\text{min}}^{\rm (BR)}$,
the black ring AH is surrounded by the black hole AH.

%===========<Picture>============%
%
\begin{figure}[tb]
\centering
{
\includegraphics[width=0.5\textwidth]{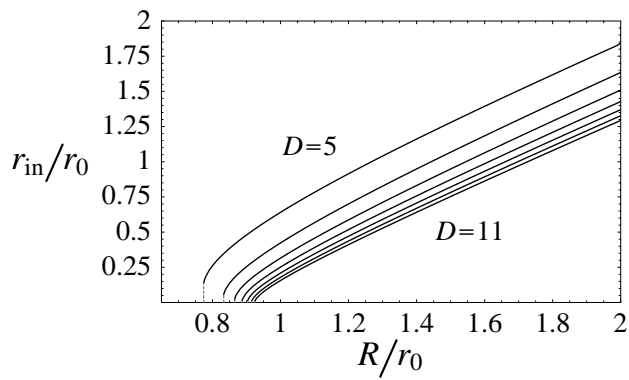}
}
\caption{ The relation between the ring radius $R$
and the inner radius of $C_{\rm BR}$,
i.e. $r_{\rm in}:=R-f(\pi)$, for $D=5$--$11$. 
The value of $dr_{\rm in}/dR$ becomes $\infty$ 
at $R=R_{\rm min}^{\rm (BR)}$. }
\label{innerradius}
\end{figure}
%
%=================================%

Let us look at the behavior of the radius of the $S^1$ circle.
We define the ``inner radius'' of  $C_{\rm BR}$ by
$r_{\rm in}:=R-f(\pi)$, which
represents the radius of the inner side
of $C_{\rm BR}$ (i.e., $\xi=\pi$). 
In Fig.~\ref{innerradius},
we plot $r_{\rm in}$ as a function of $R$ for each $D$. 
The inner radius $r_{\rm in}$
is a monotonically increasing function of $R$, 
and its gradient diverges at $R=R_{\text{min}}^{\rm (BR)}$. 
For $D\ge 7$, the value of
$r_{\rm in}$ is very small at $R=R_{\text{min}}^{\rm (BR)}$
and thus the inner side of the $C_{\rm BR}$ almost touches
as shown in Fig.~\ref{blackringAHs}.
The values of $R_{\text{min}}^{\rm (BR)}$ 
are summarized in Table~\ref{table1}.
$R_{\text{min}}^{\rm (BR)}$
is slightly less than $R_{\text{max}}^{\rm (BH)}$.
This indicates that at least one of the two AHs forms
in this system for arbitrary values of $R$. 

%
%======================================%
%<<<<<<<<<< subsection 4.3 >>>>>>>>>>>>%
%======================================%
%
\subsection{Trapped energy}

%===========<Picture>============%
%
\begin{figure}[tb]
\centering
{
\includegraphics[width=0.4\textwidth]{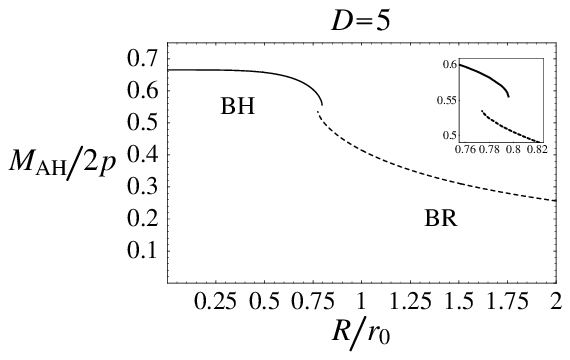}\hspace{5mm}
\includegraphics[width=0.4\textwidth]{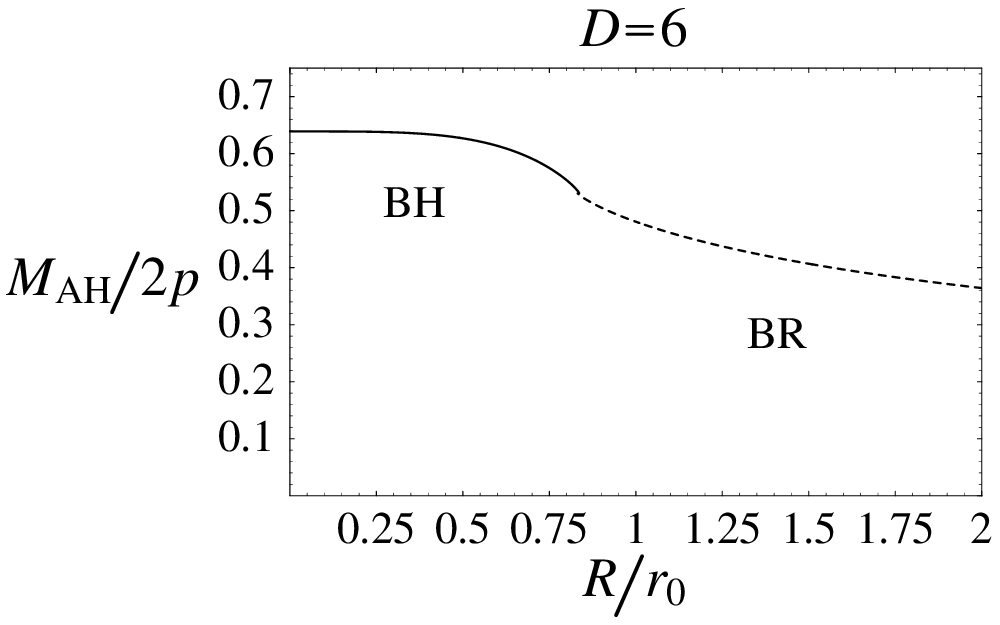}\\
\includegraphics[width=0.4\textwidth]{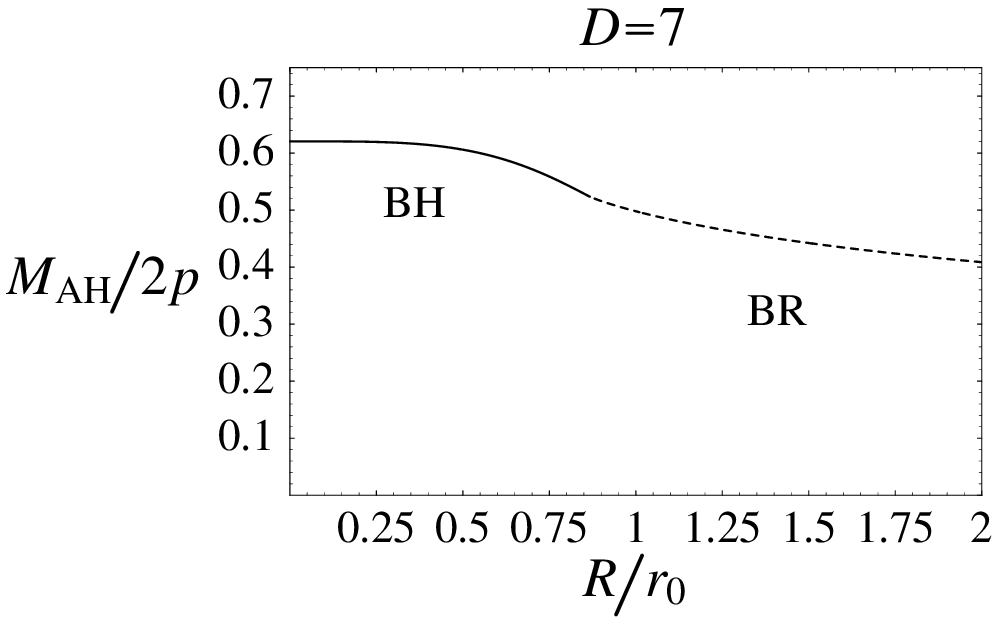}\hspace{5mm}
\includegraphics[width=0.4\textwidth]{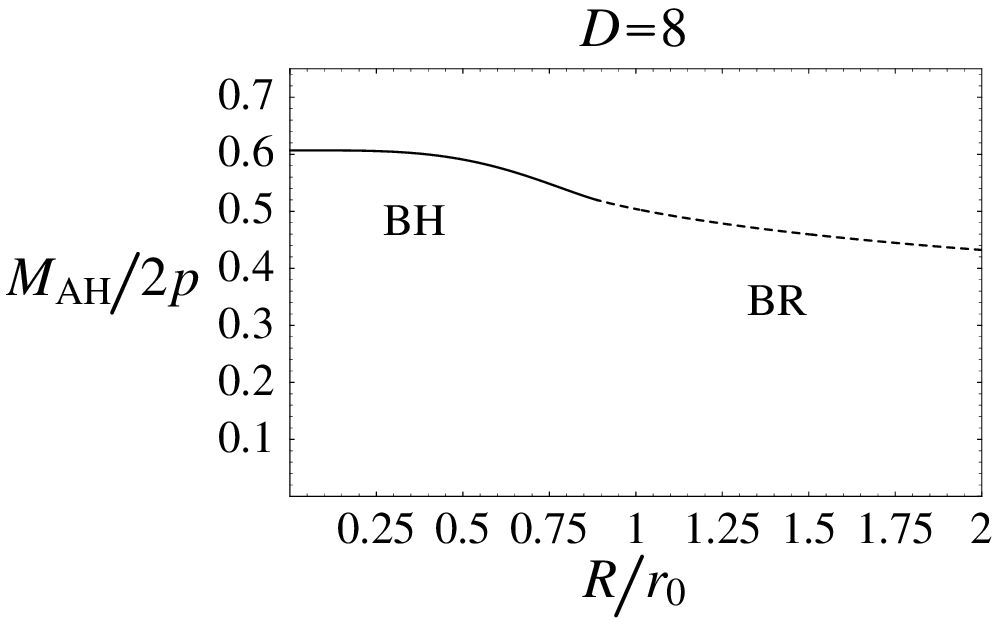}
}
\caption{ The relation between the ring radius $R$
and the trapped energy $M_{\rm AH}/2p$ by the black hole AH (solid lines)
and the black ring AH (dashed lines) for $D=5$--$8$. }
\label{MAH-vs-R}
\end{figure}
%
%=================================%

Since the event horizon is located outside of the AH, the area of
the AH $A_{D-2}$ gives the lower bound on the area of the
produced black hole. Therefore the following quantity 
%===========<Equation>============%
%
\begin{equation}
M_{\rm AH}:=\frac{(D-2)\Omega_{D-2}}{16\pi G}
\left(\frac{A_{D-2}}{\Omega_{D-2}}\right)^{(D-3)/(D-2)}
\end{equation}
%
%=================================%
provides the lower bound on the mass of the produced black hole
(or the black ring) and indicates the amount of the energy
trapped by the AH.

Figure~\ref{MAH-vs-R} shows the values of $M_{\rm AH}$ of the black hole AH
and the black ring AH as functions of $R$. The value of $M_{\rm AH}$
of the black hole AH decreases as $R$ is increased, but 
the black hole AH traps more than 50\% of the system energy
for all $0\le R\le R_{\rm max}^{\rm (BH)}$.
In the range $R_{\rm min}^{\rm (BR)}\le R\le R_{\rm max}^{\rm (BH)}$,
$M_{\rm AH}$ has two values since both the black hole AH and the black ring
AH exist. The two values of $M_{\rm AH}$ are different,
but the difference is very small for $D\ge 6$.
The value of $M_{\rm AH}$ of the black ring AH becomes
small as $R$ is increased.
For large $R$, $M_{\rm AH}/2p$ 
becomes proportional to $R^{\frac{-(D-3)}{(D-2)(D-4)}}$
since $A_{D-2}$ asymptotes to $4\pi\Omega_{D-4}R\rho_h^{D-4}$.

%
%======================================%
%<<<<<<<<<< subsection 4.4 >>>>>>>>>>>>%
%======================================%
%
\subsection{Interpretation}

%===========<Table>============%
%
\begin{table}[tb]
\caption{ The values of $R_{\text{max}}^{\rm (BH)}/r_0$,  
$R_{\text{min}}^{\rm (BR)}/r_0$ and $\beta_D^{1/(D-3)}$ for $D=5,..., 11$.
The black hole AH and the black ring AH exist for $R\le R_{\text{max}}^{\rm (BH)}$
and $R\ge R_{\text{min}}^{\rm (BR)}$, respectively. 
In all dimensions, the transition from a black hole AH to a black ring AH
occurs in a range $\beta_D^{1/(D-3)}<R/r_0<1$.} 
\begin{ruledtabular}
\begin{tabular}{c|ccccccc}
  $D$ & 5 & 6 & 7 & 8 & 9 & 10 & 11 \\
  \hline 
 $R_{\rm max}^{\rm (BH)}/r_0$ & $0.796$ & $0.836$ & $0.867$ & $0.889$ & $0.905$ & $0.917$ & $0.927$
 \\
 $R_{\rm min}^{\rm (BR)}/r_0$ & $0.774$ & $0.832$ & $0.865$ & $0.887$ & $0.903$ & $0.915$ & $0.924$
 \\
 $\beta_D^{1/(D-3)}$ & $0.564$ & $0.630$ & $0.679$ & $0.715$ & $0.744$ & $0.767$ & $0.786$
\end{tabular}
\end{ruledtabular}
\label{table1}
\end{table}
%
%=================================%

Let us briefly discuss the physical interpretation
of the obtained results. As the value of $R$ is increased,
the ``transition'' from the black hole AH to the black ring AH
occurs at $R_{\rm min}^{\rm (BR)} \le R\le R_{\rm max}^{\rm (BH)}$.
One of the plausible interpretations for the values of the transition radius is 
as follows. For $R\gtrsim 1$, the formation of the black hole AH
cannot be expected since the incoming strings have characteristic
scales larger than the gravitational radius of the system.
On the other hand, if $R$ is less than the value of 
$\rho_h$ defined in Eq.~\eqref{rhoH},
the black ring AH cannot be expected to form because the
radius of the $S^1$ circle is smaller than the radius of the $S^{D-4}$
sphere of $C_{\rm BR}$. 
Since $R=\rho_h$ is equivalent to $R=\beta_D^{1/(D-3)}$,
the transition from a black hole AH to a black ring AH
is expected to occur at $\beta_D^{1/(D-3)}<R<1$.
The values of $\beta_D^{1/(D-3)}$ are summarized in Table~\ref{table1}.
We can confirm that 
both $R_{\rm min}^{\rm (BR)}$ and $R_{\rm max}^{\rm (BH)}$
are larger than $\beta_D^{1/(D-3)}$ and smaller than $1$
for all $D$.

We also point out the similarity between our study and
that of \cite{IN02}. In that paper, several momentarily static 
initial data sets in five-dimensional spacetimes were investigated.
One of the studied system is ring-shaped matter
distribution, for which the black hole AH and the black ring 
AH form for small and large values of the ring radius, respectively.
Their interpretation is that for large ring radius, the gravitational
field in the transverse direction of the ring is four dimensional,
and thus the black ring AH can form. 
We consider that the same argument holds for our system,
since the behavior of the shock potential in the transverse direction
of the $D$-dimensional string is  similar to that
of the $(D-1)$-dimensional Aichelburg-Sexl particle,
as found by comparing Eqs.~\eqref{AS-potential} and \eqref{potential-largeR}.

%
%======================================%
%<<<<<<<<<<<< SECTION IV  >>>>>>>>>>>>>>%
%======================================%
%
\section{Discussion}

In this paper, we studied the collision of high-energy closed strings.
The positions of two closed strings were assumed to coincide
at the instant of the collision. In this setup, we found that
the black hole AH forms for small ring radius $R$
and the black ring AH forms for large ring radius $R$.
Therefore, the collision of high-energy closed strings
with large radius will lead to the formation of the black ring.

%===========<Picture>============%
%
\begin{figure}[tb]
\centering
{
\includegraphics[width=0.4\textwidth]{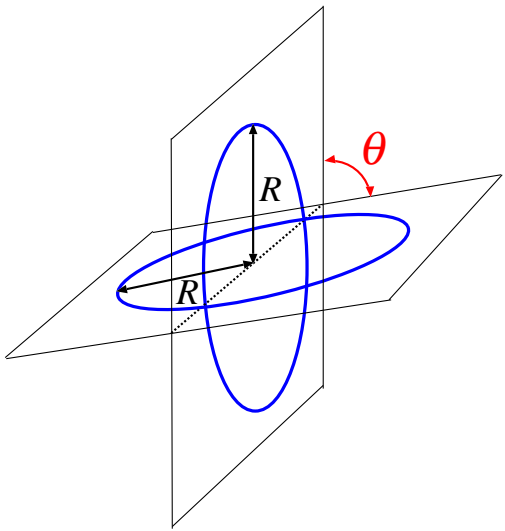}
}
\caption{ The example of positions of two closed strings 
at the instant of the head-on collision in five-dimensional case. 
The direction of the motion is suppressed. Because the shock
is three dimensional, there is a nonzero angle $\theta$ between
the two planes on which the strings exist.}
\label{generalcase}
\end{figure}
%
%=================================%

Because the setup of the collision in this paper is quite limited,
let us discuss what happens in a more general setup.
For simplicity, let us assume that the radii of two strings be the same.
In general collisions, we have to take account of the impact parameter
$b$. Here, $b$ is defined as the distance of two centers
of incoming strings when two shocks collide.
Furthermore, even in the head-on collision $b=0$,
the orientations of two strings do not coincide
at the instant of the collision. Figure \ref{generalcase}
depicts the example in the five-dimensional case.
Because the shock is three dimensional, the two planes on which 
two strings exist cross each other at angles $\theta$.
Similarly, in the $D$-dimensional case, the relative orientation
of two strings will be specified by angles $\theta_1,...,\theta_{D-4}$.

Let us first discuss the black hole AH. In the head-on collision $b=0$,
it is expected that the main factor for the AH formation
is the ring radius, because the discussion in Sec.~IVD
holds for any angles $\theta_i$.
Namely, the maximum ring radius $R_{\rm max}$
for the formation of the black hole AH would 
be about $r_0$ and would weakly depend on $\theta_i$.
As the impact parameter $b$ is increased, 
$R_{\rm max}$ will become small
and eventually go to zero.
This happens at $b\simeq r_0$, since 
in the Aichelburg-Sexl particle case ($R=0$),
the maximal impact parameter for the AH formation is 
$b_{\rm max}\simeq r_0$
as demonstrated in \cite{YN03, YR05}. 
Therefore, we expect that the condition
for the black hole formation will be given by $b\lesssim r_0-R$.
This is the same criterion as the one discussed in \cite{KV02}.

Next we discuss the black ring AH.
The condition for the formation of the black ring AH 
will strongly depend on the angles $\theta_i$ even in 
the head-on collision cases $b=0$, since
the distance between the two strings should be less than
$\rho_h$. Namely, in the five-dimensional case, 
the angle $\theta$ should satisfy $|\theta|\lesssim \Delta\theta:=
\rho_h/R\sim R^{-2}$. Similarly, in the $D$-dimensional case,
all the angles $\theta_i$ should satisfy $|\theta_i|\lesssim \Delta\theta:=
\rho_h/R\sim R^{-(D-3)/(D-4)}$. 
Therefore, even in the head-on collision cases,
the black ring formation is quite difficult.
If we cannot control the orientations of the incoming
strings, the black ring will form only with the probability 
$\sim \Delta\theta^{D-4}\sim R^{-(D-3)}$.
 Furthermore,
the impact parameter should be smaller than the
the radius of the $S^{D-4}$ sphere of $C_{\rm BR}$,
i.e. $b\lesssim \rho_h\sim R^{-1/(D-4)}$, which is much smaller than $r_0$.

Finally we discuss some implication for
the trans-Planckian collisions of fundamental strings
in the scenarios of large extra dimensions.
Let us consider the collisions of closed strings with 
length $\lambda_s=2\pi R$ and assume that our 
system can describe these processes.
Under this assumption, we find that
$R\lesssim r_h(2p)$ is necessary for the black hole formation.
For $R\gtrsim r_h(2p)$,  the black ring formation
would be expected instead, with the probability discussed above. 
However, we point out here that an additional
condition is required for the production
of a classical black ring. Because the black ring
can be a classical object only when 
the radius of the $S^{D-4}$ sphere of $C_{\rm BH}$ is 
larger than the Planck length $l_p$,
the system should satisfy $\rho_h\gtrsim l_p$.
This is rewritten as $\frac{R}{l_p}\lesssim\left(\frac{r_h(2p)}{l_p}\right)^{D-3}$.
Therefore, the parameter range of $R$ for the black ring formation
is restricted also from above.

Since we studied the collision of closed strings and the closed
strings do not stand for gauge particles, our results cannot
be directly applied for the phenomena in accelerators.
We are planning to generalize our result to the collision of 
high-energy open strings.

%
%======================================%
%<<<<<<<<< Acknowledgements >>>>>>>>>>>%
%======================================%
%\baselineskip25pt
%
\acknowledgments

HY thanks the Killam Trust for financial support.  
The work of TS was supported by Grant-in-Aid for
Scientific Research from Ministry of Education, Science, Sports
and Culture of Japan (No.~13135208, No.~14102004, No.~17740136 and
No.~17340075), the Japan-U.K. and Japan-France Research
Cooperative Program. 

\appendix
%
%======================================%
%<<<<<<<<<<<< Appendix A  >>>>>>>>>>>>>>%
%======================================%
%
\section{Calculation of the shock potential}  

In this section, we show the formulas of $I_D$
defined in Eq.~\eqref{integral}. We begin with even $D$ values.
For $D=6$, 
\begin{equation}
I_6=\frac{2\pi}{\sqrt{a^2-b^2}}.
\end{equation}
By taking derivatives of this formula with respect to $a$, 
we obtain
\begin{equation}
I_8=\frac{2\pi a}{(a^2-b^2)^{3/2}};
\end{equation}
\begin{equation}
I_{10}=\pi\frac{2a^2+b^2}{(a^2-b^2)^{5/2}}. 
\end{equation}

Next we show the cases of odd $D$ values.
In these cases, the integrals are expressed 
in terms of the complete elliptic integrals of the first and second kinds:
\begin{equation}
K(k):=\int_0^{\pi/2}\frac{d\theta}{\sqrt{1-k^2\sin^2\theta}};
\end{equation}
\begin{equation}
E(k):=\int_0^{\pi/2}d\theta\sqrt{1-k^2\sin^2\theta}.
\end{equation}
$I_5$ is found to be
\begin{equation}
I_5=\frac{4}{\sqrt{a+b}}K(k)
\end{equation}
with $k:=\sqrt{2b/(a+b)}$. 
By taking derivatives of this formula with respect to $a$, we obtain
\begin{equation}
I_7=\frac{4}{(a-b)\sqrt{a+b}}E(k); 
\end{equation}
\begin{equation}
I_9=\frac{4}{3(a-b)^2(a+b)^{3/2}}\left[4aE(k)-(a-b)K(k)\right];  
\end{equation}
\begin{equation}
I_{11}=
\frac{4}{15(a-b)^3(a+b)^{5/2}}\left[(23a^2+9b^2)E(k)-8a(a-b)K(k)\right]. 
\end{equation}

%---------   References   ---------%

%---------   References   ---------%

\end{document}